\documentclass[12pt]{article}
\usepackage{amssymb,amsmath}
\usepackage{epsfig}

\def\appendix#1{
  \addtocounter{section}{1}
 \setcounter{equation}{0}
  \renewcommand{\thesection}{\Alph{section}}
 \section*{Appendix \thesection\protect\indent \parbox[t]{11.715cm} {#1}}
  \addcontentsline{toc}{section}{Appendix \thesection\ \ \ #1}
  }

\newcommand{\newsection}{
\setcounter{equation}{0}
\section}

\newcommand{\eq}[1]{\begin{equation} #1 \end{equation}}
\newcommand{\ar}[1]{\begin{eqnarray} #1 \end{eqnarray}}
\newcommand{\tr}{\mathop{\mathrm{tr}}\nolimits}

\def\const{{\rm const}}
\def\e{{\,\rm e}\,}
\def\d{\partial}
\def\D{\delta}

\newcommand{\br}[1]{\left( #1 \right)}
\newcommand{\vev}[1]{\left\langle #1 \right\rangle}
\newcommand{\rf}[1]{(\ref{#1})}
\newcommand{\non}{\nonumber \\*}
\def\ra{\rightarrow}
\def\lra{\longrightarrow}
\def\ll2{\ll}

\def\N{${\cal N}=4$ }
\def\th{\theta}

\def\st{\sqrt{\lambda}}

\def\cir{{\rm circle}}
\def\O{{\cal O}}
\def\ll{\lambda}
\def\by{{\bf Y}}
\def\ads{$AdS_5\times S^5$ }
\def\oj{\O_J}
\def\cj{c_J}
\def\ap{\sqrt{\lambda}}
\newcommand{\cl}[1]{#1_{\rm cl}}
\def\sg{\sigma}
\def\zf{\zeta^4}
\def\tz{\tilde{\zeta}}
\def\lt{\lim_{\sigma'\rightarrow\sigma}}
\def\ep{\varepsilon}

\title{
\hfill{\small UUITP-10/02}\\
\hfill{\small ITEP-TH-41/02}
\\~\\
Open string fluctuations in $AdS_5\times S^5$ 
\\[0.3cm]
and operators with large R-charge}

\author{K. Zarembo\thanks{{\tt Konstantin.Zarembo@teorfys.uu.se}. Also at
ITEP, B.~Cheremushkinskaya 25, 117259 Moscow, Russia}
\\~~\\
Department of Theoretical Physics\\ 
Uppsala University \\
Box 803, SE-751 08 Uppsala, Sweden 
}

\begin{document}           

\maketitle 
               
\abstract{A semiclassical string description is given for
correlators of Wilson loops with local operators in \N
SYM theory in the regime when operators 
carry parametrically large R-charge. 
The OPE coefficients of the circular Wilson loop 
in chiral primary operators
are computed
to all orders in the $\alpha'$ expansion in
\ads string theory. The results agree with field-theory
predictions.
}


\newsection{Introduction}

The string states with large quantum numbers are semiclassical
and consequently are simple even in highly curved backgrounds. 
This fact has proven extremely useful in the context of
the holographic duality between type IIB 
strings in \ads and \N supersymmetric Yang-Mills (SYM) theory
\cite{review}
and has lead to an elegant string description
of sectors of large spin or large R-charge in SYM theory
\cite{BMN,GKP}. 
This description
goes beyond the strong-coupling supergravity limit and 
can be compared to perturbative SYM theory at weak coupling
\cite{BMN,GKP,Kristjansen:2002bb,Gross:2002su,
Constable:2002hw,Santambrogio:2002sb}. 
The string states with large
R-charge correspond to excitations of closed point-like
strings that carry
large angular momentum in $S^5$ \cite{GKP}. 
The purpose of the present paper is
to study how these states couple to the open string sector, which
is dual to Wilson loops in the SYM theory \cite{MAL,Rey,SZ}. 
Though open strings can also rotate in $S^5$,
their angular momentum  never becomes large \cite{Tseytlin:2002tr}.
To make contact with the limit of large R-charge,
  I will exploit the fact that any Wilson loop contains, as a composite
operator, an admixture of states with any quantum numbers. 
It is natural to expect that the expansion 
coefficients of a Wilson loop in operators with large R-charge 
will have some kind of semiclassical string description.

The standard OPE of a Wilson loop
 \cite{Shifman1980:ui,Berenstein:1999ij} is defined as
\begin{equation} 
W(C)= \langle W(C)\rangle\sum
c_A R^{\Delta_A} {\cal O}_A(0), 
\label{expansion}
\end{equation} 
where ${\cal O}_A(0)$ is an operator evaluated at the center of
the loop, $\Delta_A$ is the conformal dimension of ${\cal O}_A(x)$,
and $R$ is the radius of the loop ($R$ will be set to 1
in what follows). Dimensionless
coefficients $c_A$ depend on the shape of the contour $C$
and, in the large-$N$ limit, on the 't~Hooft coupling
$\ll=g^2_{YM}N$. To start with, I will consider a specific case of the
circular Wilson loop and of its expansion coefficients
in the chiral primary operators. 
Then I will give some more general results in 
sec.~\ref{stringpert}. The operators of interest are
\eq{\label{oj}
\O_J=\frac{(2\pi)^J}{\sqrt{J}\,\ll^{J/2}}\,\tr Z^J,
}
where $Z=\Phi_1+i\Phi_2$. These operators have R-charge $J$ 
and dimension $\Delta=J$, and play an important role
in the planar-wave limit of the AdS/CFT correspondence 
\cite{BMN}. The reason to choose the circular loop is two-fold.
First, there is a lot of explicit result for the circular loop
on the string side of AdS/CFT duality 
\cite{Berenstein:1999ij,DGO,DGT}. Second,  
the invariance of the circular loop operator
under certain superconformal transformations \cite{Bianchi:2002gz}
partially protects it 
from gaining quantum corrections \cite{Drukker:2000rr}. 
An expectation value of
the circular loop \cite{Erickson:2000af,Drukker:2000rr} 
and some of its OPE coefficients \cite{Semenoff:2001xp}
can be exactly calculated by resumming Feynman diagrams that
survive supersymmetry cancellations. In particular, weights
with which operators \rf{oj} appear in the circular Wilson loop 
are known \cite{Semenoff:2001xp}:
\eq{\label{cj}
c_J(\ll)=\frac{1}{2}\,\sqrt{J\ll}\,\,\frac{I_J\br{\st}}{I_1\br{\st}}\,.
}
Here, $I_J(x)$ are modified Bessel functions. Some of their
properties are listed in appendix~A.
This expression is believed to be exact in the large-$N$
limit.

In string theory, OPE coefficients of a Wilson loop measure
an overlap of the boundary state associated with the loop with the closed
string state associated with the local operator. The overlap can be 
computed in the supergravity approximation \cite{Berenstein:1999ij}, 
which corresponds to taking
the limit of large $\ll$ in \rf{cj}. Further expansion 
in $1/\ap$ should be equivalent to the semiclassical (or $\alpha'$)
expansion in string theory. The purpose of the present paper is
to  go beyond the supergravity limit directly
in the \ads string theory and
to extend the semiclassical approximation for OPE coefficients
to all orders in $\alpha'$
for operators with large R-charge $J$.
Before going into details, let me illustrate on a simple example how
large quantum numbers can modify a semiclassical expansion. 

\newsection{A toy model}

\begin{figure}[h]
\begin{center}
\epsfxsize=9cm
\epsfbox{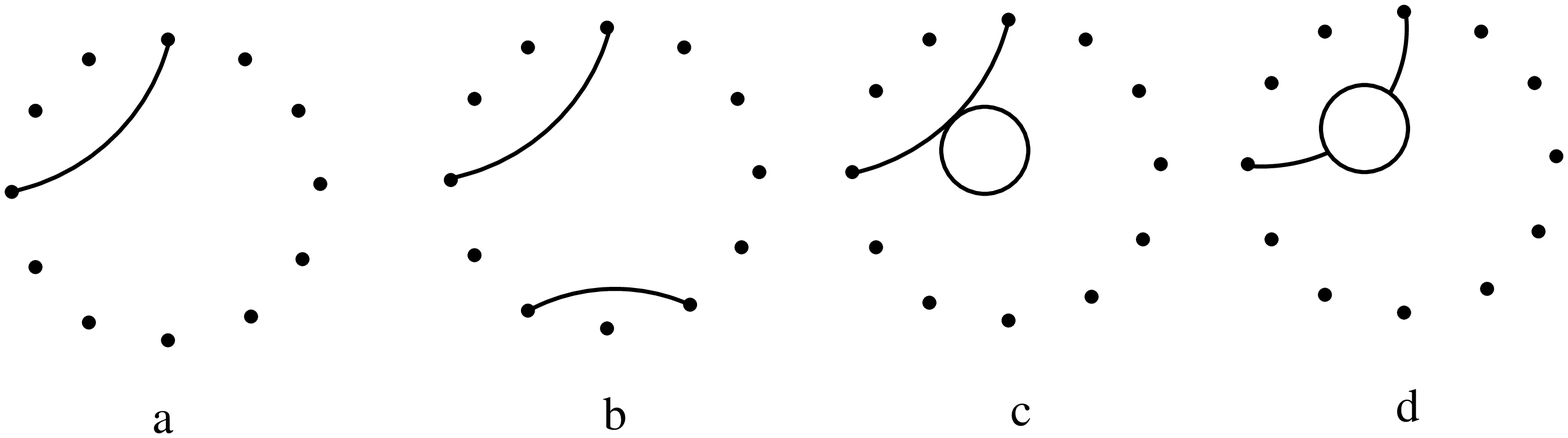}
\end{center}
\caption[x]{\small The semiclassical expansion of $\protect\vev{\phi^n}$
to order $O(\hbar^2)$.
The first two diagrams are of order $\hbar n^2$ and
$\hbar^2n^4 $, respectively. The loop corrections (c) and (d) are of order
$\hbar^2n^2$ and are down by a factor of $n^2$ compared to (a) and (b).}
\label{diagr}
\end{figure}

Consider an integral
\eq{\label{toy}
\vev{\phi^n}=\frac{\int D\phi\,\phi^n\e^{-\frac{1}{\hbar}\,S(\phi)}}
{\int D\phi\,\e^{-\frac{1}{\hbar}\,S(\phi)}}.
}
and suppose that $\hbar$ is small. The integral is then dominated
by a saddle point of the action: $S'(\cl{\phi})=0$. The semiclassical
expansion is generated by shifting
$\phi=\cl{\phi}+\sqrt{\hbar}\xi$ and expanding in $\xi$:
\eq{\label{sem}
\vev{\phi^n}=\cl{\phi}^n+\frac{n(n-1)}{2!}\,\hbar\vev{\xi^2}\cl{\phi}^{n-2}
+\frac{n(n-1)(n-2)(n-3)}{4!}\,\hbar^2\vev{\xi^4}\cl{\phi}^{n-4}+\ldots\,.
}
It is clear that  
the semiclassical expansion breaks down if $n$ is sufficiently large,
because the expansion parameter
is not really $\hbar$, but $\hbar n^2$. The semiclassical
series can be easily resummed in the double-scaling limit when $\hbar$ goes to
zero and $n$ goes to infinity with $n\sqrt{\hbar}$ held fixed. 
Simplifications occur because loops are suppressed by extra powers
of $\hbar$ and only tree diagrams survive the
double-scaling limit
(fig.~\ref{diagr}). It is easy to show that
relevant tree diagrams exponentiate:
\eq{\label{tm}
\vev{(\cl{\phi}+\sqrt{\hbar}\xi)^n}=
\cl{\phi}^n\vev{\br{1+\frac{\sqrt{\hbar n^2}}{\cl{\phi} n}\,\xi}^n}
\longrightarrow\cl{\phi}^n
\vev{\exp\br{\frac{\sqrt{\hbar n^2}}{\cl{\phi} }\,\xi}}
_{\rm Gauss.}\,.
}
The Gaussian average can be easily computed:
\eq{
\vev{\phi^n}\approx\cl{\phi}^n\exp\br{\frac{\hbar n^2}{2\cl{\phi}^2}\,D},
}
where $D=[S''(\cl{\phi})]^{-1}$ is the propagator of $\xi$.
To facilitate the double-scaling
limit, it is convenient to set $\phi=\cl{\phi}\e^\zeta$ from the start
and to treat $\zeta$ as a Gaussian fluctuation.

The parameter of the
semiclassical expansion in AdS string theory is
 $1/\sqrt{\ll}$, which plays the role of the sigma-model coupling
$\alpha'$ and
is analogous to $\hbar$ of the toy model. The counterpart of $n$ is the
R-charge $J$. The toy model suggests that $J^2/\ap$ is 
an expansion parameter of the $\alpha'$ expansion and that
the expansion 
breaks down at $J\sim\ll^{1/4}$.

Let us now expand the exact OPE coefficient \rf{cj} in powers of
$1/\ap$:
\eq{\label{semstr}
c_J(\ll)=\frac{1}{2}\,\sqrt{J\ll}\br{1-\frac{J^2-1}{2\st}
+\frac{J^4-4J^2+3}{8\ll}
+\ldots}.
}
This has precisely the same form as \rf{sem} and, indeed,
$J^2/\ap$ emerges as a parameter of the strong-coupling
expansion. 
Moreover, the series exponentiate in the double-scaling
limit of large $J$ and large 
$\ll$\footnote{A derivation is given in
appendix~A.}:
\eq{\label{dsl}
c_J(\ll)\approx\frac{1}{2}\,\sqrt{J\ll}\,\exp\br{-\frac{J^2}{2\ap}}.
}
These observations suggest that the large-$J$ limit of OPE
is described by the double-scaling limit of the $\alpha'$ expansion 
in \ads sigma model. 

There is another way to take the limit of large $n$ in the toy model.
Its counterpart in string theory closely follows Polyakov's approach
\cite{strings2002}
to string amplitudes with large angular 
momentum\footnote{I would like to thank A.~Tseytlin for
this remark.}. The idea is to treat an operator insertion $\phi^n$
on an equal footing with the action, that is, to minimize
$W(\phi)=S(\phi)-\hbar n\ln\phi$ instead of $S(\phi)$.
The expectation value then takes the form $\vev{\phi^n}\approx
\exp(-f(\hbar n)/\hbar)$, which clearly assumes that $n\sim \hbar^{-1}$.
In the AdS/CFT context, this translates into the BMN scaling 
$J\sim\ap$, which leads to the plane wave limit
of \ads geometry in the closed-string sector \cite{BMN}. 
Does the BMN limit make sense for the OPE
coefficients \rf{cj}? A simple calculation (see appendix~A) shows
that it does and that OPE coefficients indeed have 
the expected exponential
form in the limit:
\eq{\label{Blim}
\cj=\const\cdot\exp\left\{-\ap\left[1-\sqrt{j^2+1}-j\ln\br{\sqrt{j^2+1}-j}
\right]\right\},
}
where $j=J/\ap$. When $j$ goes to zero, this expression reduces to
 \rf{dsl}. Again, there is a strong evidence that the BMN limit
of OPE coefficients can be computed in the sigma model by
semiclassical techniques. 


\newsection{OPE expansion at large $J$}
\label{j4}

The Wilson loop operator which couples to strings in \ads
is defined as \cite{MAL}
\eq{\label{php} W(C)=\frac{1}{N}\,\tr{\rm P}\exp\left[\oint_C
ds\,\br{i A_\mu(x)\dot{x}^\mu +\Phi_i(x)\theta^i|\dot{x}|}\right],
} 
where $\theta_i$ is a unit 6-vector: $\theta^2=1$.
The OPE coefficients of this operator can be extracted from
the normalized two-point function
\eq{\label{corr}
\frac{\vev{W(\cir)\O_J(x)}}{\vev{W(\cir)}}
=\frac{1}{N}\,c_J(\ll)\,
\frac{1}{|x|^{2J}}\,2^{J/2}\by_J(\theta)+\ldots\,,
}
where omitted terms correspond to descendants of $\oj$ and
are suppressed by powers of $1/|x|$. A factor of $2^{J/2}$
is introduced for later convenience and $\by_J(\theta)$
is the spherical function associated with operator $\oj$:
\eq{
\by_J(\theta)=\br{\frac{\theta_1+i\theta_2}{\sqrt{2}}}^J.
} 
A convenient choice is $\theta^i=(1,0,\ldots,0)$, since
then
\eq{
2^{J/2}\by_J(1,0,\ldots,0)=1.
} 
For the two-point function \rf{corr}, the free-field approximation
is believed to be exact and gives the result \rf{cj} quoted
in the Introduction.

On the string side of the AdS/CFT correspondence, Wilson loops
are associated with open strings that end on the boundary
of \ads.
The expectation value of a Wilson loop is equal to
the sigma-model partition function with boundary conditions
set by the contour $C$ 
 and by the point $\theta_i$
in $S^5$:
\ar{\label{equality}
\vev{W(C)}&=&\int DX^\mu DY D\varTheta_i Dh_{ab} D\vartheta^{\alpha}
\,\D(\varTheta^2-1)
\non
&& \times
\exp
\left[-\frac{\sqrt{\lambda}}{4\pi}\,\int_D
d^2\sigma\,\sqrt{h}\,h^{ab}\,\left(\frac{
\partial_a X^\mu\partial_b X^\mu+\partial_a Y\partial_b Y}{Y^2}
+\d_a\varTheta_i\d_b\varTheta_i\right)
\right.
\non
&&\left.
\vphantom{
-\frac{\sqrt{\lambda}}{4\pi}\,\int
d^2\sigma\,\sqrt{h}\,h^{ab}\,\frac{
\partial_a X^\mu\partial_b X^\mu+\partial_a Y^i\partial_b Y^i}{Y^2}
} 
+{\rm fermions} \right].
\label{strpartfn}
}
Here, $\vartheta^{\alpha}$ are world-sheet fermions
and $y$ is the AdS scale\footnote{In what follows, 
the bosonic world-sheet coordinates
are denoted by capital letters: $Y$, $X^\mu$, {\it etc}.
}. 
The boundary of \ads is at $y=0$.

The chiral primary operators of R-charge $J$ are dual
to $J$th KK 
modes on $S^5$ of a particular supergravity field.
The two-point function \rf{corr} describes emission of one
such mode by a source at point $x$ on the boundary of 
$AdS_5$, its propagation in the bulk, and subsequent
absorption by the string world sheet. 
When $|x|$ is large, the supergraviton propagator
associated with dimension-$J$ operator behaves as $y^J/|x|^{2J}$.
The denominator gives rise to an overall factor of $1/|x|^{2J}$ 
in the correlation
function, and $Y^J$ can be regarded as a part of the vertex
operator which couples the supergravity mode to the world sheet. 
The OPE coefficient is essentially a one-point function
of this vertex operator. In other words, it is an overlap of the
boundary state associated with the Wilson loop with the graviton
state associated with the local operator. It is easy to guess which terms 
in the vertex operator are responsible for
enhancement of the $\alpha'$  expansion at large $J$.
Those are the factor of $Y^J$, which comes from the supergraviton
propagator, and $\by_J(\varTheta)$, which is the wave function 
of $J$th KK mode on $S^5$. These simple arguments determine the
vertex operator up to a coefficient.
A more precise form of the
vertex operator, including all coefficients, was derived in
\cite{Berenstein:1999ij}:
\eq{\label{ex}
c_J=2^{J/2}\sqrt{J\ll}\,\frac{J+1}{4\pi}
\vev{\int d^2\sigma\,\sqrt{h}\,Y^{J+2}\by_J(\varTheta)
}.
}
The averaging over string fluctuations here is determined by the
partition function \rf{strpartfn}. The above expression still
contains some approximations \cite{Berenstein:1999ij}: an exact 
vertex operator is more complicated, in particular, it contains
world-sheet derivatives and fermions, but those appear only
in the sigma-model loops, 
so \rf{ex} is sufficient for semiclassical calculations.

It is convenient to parameterize the five-sphere by two angles,
$\psi$ and $\varphi$, and a unit three-vector ${\bf n}$:
\eq{
\th=(\cos\psi\cos\varphi,\cos\psi\sin\varphi,\sin\psi\,{\bf n}),
~~~~~{\bf n}\in S^3.
}
In this parameterization,
the spherical functions read
\eq{
\by_J(\psi,\varphi,{\bf n})=2^{-J/2}(\cos\psi)^J\e^{iJ\varphi},
}
and the OPE coefficients become:
\eq{\label{main}
c_J=\sqrt{J\ll}\,\frac{J+1}{4\pi}
\vev{\int d^2\sigma\,\sqrt{h}\,Y^{J+2}(\cos\varPsi)^J\e^{iJ\varPhi}
}.
}

At large $\ll$, the  path integral is dominated
by a saddle point, a minimal surface with the fixed boundary.
For the circle of unit radius, the classical solution
is \cite{Berenstein:1999ij,DGO}
\eq{\label{mins}
\cl{Y}=z,~~~~~\cl{X^1}=r\cos\phi,~~~~~\cl{X^2}=r\sin\phi,
~~~~~\cl{\varPsi}=0,~~~~~\cl{\varPhi}=0,~~~~~r=\sqrt{1-z^2}.
}
In the coordinates $(z,\phi)$, the induced metric takes the form:
\eq{\label{metr}
g_{ab}d\sg^a d\sg^b=\frac{1}{r^2z^2}\,dz^2+\frac{r^2}{z^2}\,d\phi^2.
}
The substitution of the classical solution into the expectation value
\rf{main} gives $\cj=\sqrt{J\ll}/2$ \cite{Berenstein:1999ij}, 
which coincides with the
strong-coupling limit of the OPE coefficient computed in SYM theory
\rf{cj}. This result can be improved by taking into account
string fluctuations. In the scaling limit $\ll\ra\infty$, $J\ra\infty$,
$J^2/\ap$-fixed, we can repeat the same steps as in the toy model
of the previous section, since the expectation value we need to
calculate has essentially the same structure as \rf{toy}.

The first step is to expand the action of the sigma model around
the classical solution. We only need the quadratic part of the action, 
but even that is not so simple, because various
fluctuations mix. Diagonalization
and expansion of fluctuations in normal modes is
a non-trivial problem. For the circular loop, it was solved in 
sec.~5 of \cite{DGT}. The results which will be necessary to calculate 
the OPE coefficients are shortly summarized below.
The metric fluctuations are eliminated by imposing background 
conformal gauge:
$h_{ab}=\e^\chi g_{ab}$, where $\chi$ is an arbitrary conformal factor. 
In the notations of \cite{DGT}, the world-sheet fields are
expanded around the classical solution as follows: 
\eq{\label{expand}
Y=z\e^{\zf},~~~~~X^1=(r+z\zeta^1)\cos\phi-z\zeta^0\sin\phi,
~~~~~X^2=(r+z\zeta^1)\sin\phi+z\zeta^0\sin\phi.
}
We should treat $\zeta^a$ and all other fields, including
$\varPhi$ and $\varPsi$, as Gaussian fluctuations. 
The radial fluctuations $\zeta^1$, $\zeta^4$ mix and are expanded in
normal modes as
\ar{\label{mix}
\zf&=&z\tz^4-r\tz^1,
\\*
\zeta^1&=&r\tz^1+z\tz^4.
}
The relevant part of the action for normal modes is
very simple \cite{DGT}:
\eq{\label{smf}
S=\frac{\ap}{4\pi}\int d^2\sg\,\sqrt{g}\,g^{ab}
\left[\d_a\varPhi\d_b\varPhi+\d_a\tz^4\d_b\tz^4+2(\tz^4)^2
+\ldots\right].
}

Substituting \rf{expand} into the
correlation function \rf{main}, we get:
 \eq{\label{intrm}
\cj=\sqrt{J\ll}\,\frac{J+1}{4\pi}
\int_0^{2\pi} d\phi\,\int_0^1 dz\,
z^J
\vev{
\e^{J(\zf+i\varPhi)}}_{\rm Gauss.}
}
The fluctuations of $\varPsi$ do not contribute to the leading order in $J$,
because expansion of $(\cos\varPsi)^J$  in $\varPsi$ starts from the quadratic 
term and leads to $O(J/\ap)$ correction to the OPE coefficient, 
which vanishes
in the scaling limit we consider. 
The Gaussian average is defined by the action \rf{smf} and is
expressed in terms of standard scalar propagators on the 
minimal surface:
\eq{
\br{-\frac{1}{\sqrt{g}}\,\d_a g^{ab}\sqrt{g}\,\d_b+m^2}
G_{m^2}(\sg,\sg')=\frac{1}{\sqrt{g}}\,\D(\sg,\sg').
}
Doing the Gaussian integral, we get
\eq{
\vev{\e^{J\br{\zf(\sg)+i\varPhi(\sg)}}}_{\rm Gauss.}
=\lt\exp \left[\frac{\pi J^2}{\ap}\br{z^2 G_2(\sg,\sg')+r^2\tilde{G}(\sg,\sg')
-G_0(\sg,\sg')}\right],
} 
where $\tilde{G}(\sg,\sg')$ is the propagator of $\tz^1$,
which will drop out at the end of the calculation.
Individual propagators are singular at coinciding points, but
the singularities cancel between the $AdS_5$ and the $S^5$
contributions, so the one-point function of the vertex
operator is finite. It remains to integrate it over the world sheet.
The integration is, in fact, trivial because:
\eq{
(J+1)\int_0^1 dz\,z^J f(z)=f(1)+O(1/J)
} 
for any smooth function. Because of the $z^J$ factor, 
the main contribution comes 
from the point on the world sheet
which is farthest from the boundary of $AdS_5$. Since
$r\equiv\sqrt{1-z^2}=0$ at $z=1$, the propagator $\tilde{G}(\sg,\sg')$
drops out from the final result:
\eq{\label{gen}
c_J=\frac{1}{2}\,\sqrt{J\ll}\exp\br{-\frac{A\pi J^2}{\ap}},
}
where
\eq{
A=\lt\br{G_0(\sg,\sg')-G_2(\sg,\sg')}.
}
An agreement with the SYM prediction \rf{dsl} requires
$A=1/2\pi$.  A simple calculation, which is deferred to appendix~B, gives
precisely the right value for $A$.

Though calculations have been done only for the circle, 
an enhancement of the semiclassical
expansion at $J\sim\ll^{1/4}$ must be a generic feature. 
It is not hard to see that  OPE coefficients of
an arbitrary Wilson loop behave as
$$
a_J\sqrt{\ll}\exp\br{-\frac{A\pi J^2}{\ap}}
$$ 
at large $J$ and large $\ll$. The constants 
$a_J$ and $A$ depend on the shape of the loop and on a particular
choice of local operators. $A$ must be positive,
because operators of large R-charge must be exponentially suppressed
in the OPE expansion,
according to quite general arguments \cite{Zarembo:1999bu,SZ}. 

\newsection{BMN limit of OPE coefficients}

When $J\sim\ap$, the exponent in the vertex operator
and the sigma-model action are of the same order. 
The integral over string world-sheets is then dominated
by a solution of classical equations of motion with the
source term which comes from the
operator insertion \cite{strings2002}. 
It is convenient to use an exponential
parameterization of the AdS radial coordinate: $y=\e^{-p}$.
Then the string action takes the following form
in the conformal gauge:
\ar{\label{acac}
S&=&\frac{\ap}{4\pi}\int d^2\sg'\,\left[(\d P)^2+\e^{2P}(\d X^\mu)^2
+(\d\varPsi)^2+\cos^2\varPsi\,(\d\varPhi)^2\right]
\non
&&
+JP(\sg)-iJ\varPhi(\sg)-J\ln\cos\varPsi(\sg).
}
Not only the path integral over string world sheets, but also
the integral over the position of operator insertion is
semiclassical, so the action should be minimized with
respect to the point of operator insertion $\sg$, too. 

Before going to the general case, let us first 
consider the solution of equations
of motion at $J=0$. To find it, we should transform
 \rf{mins} to the conformal gauge:
\eq{\label{mins1}
\cl{Y}=\tanh\tau,~~~~~\cl{X}^1=\frac{\cos\phi}{\cosh\tau},~~~~~
\cl{X}^2=\frac{\sin\phi}{\cosh\tau}.
}
The induced metric on the minimal
surface is
conformal to the unit metric
in the $(\tau,\phi)$ coordinates: $ds^2=(d\tau^2+d\phi^2)/\sinh^2\tau$.
The world sheet is a semi-infinite cylinder, which is mapped to
the hemi-sphere in the target space. The center of symmetry on the hemi-sphere
corresponds to $\tau=\infty$. 

Returning to the general case of an arbitrary
angular momentum, we can notice that the minimum of the action
in the position of operator insertion is reached at the most 
symmetric point, when the vertex 
operator is inserted at $\tau=\infty$. Consequently, the minimal surface
possesses rotational symmetry at any $J$, not only
at $J\neq 0$, and we can take
the following ansatz for the classical string world sheet:
\eq{
Y=\e^{-P(\tau)},~~~~~X^1=R(\tau)\cos\phi,~~~~~X^2=R(\tau)\sin\phi,
~~~~~\varPhi=\varPhi(\tau).
}
The azimuthal angle on $S^5$ can be put to zero,
because $\varPsi=0$ satisfies equations of motion even
in the presence of the source. Substituting the ansatz into the
action, we get
\eq{\label{ac}
S=\ap\left\{\frac{1}{2}\int_0^\infty d\tau\,
\left[\dot{P}^2+\e^{2P}\br{\dot{R}^2+R^2}+\dot{\varPhi}^2\right]
+jP(\infty)- ij\varPhi(\infty)\right\},
}
where the dot denotes the derivative in $\tau$.
The source term in the action sets boundary conditions at infinity.
To understand what exactly are these boundary conditions,
it is convenient to map
the cylinder to a disc by changing coordinates from 
$\tau$ to $\rho=\e^{-\tau}$.
The world-sheet metric becomes 
$ds^2\propto(d\rho^2+\rho^2 d\phi^2)$. If a free 
field with the action normalized as in \rf{acac} 
interacts with a source of strength $J$ localized at
$\rho=0$, it asymptotes
$-j\ln\rho$ at the center of the disc. The logarithmic 
singularity on the disc translates into the
linear ($\sim j\tau$) growth at infinity on the cylinder.
Therefore, the boundary conditions are
\eq{
R(\tau)\ra 0,~~~~~P(\tau)\ra -j\tau,~~~~~\varPhi(\tau)\ra ij\tau
~~~~~({\rm at}~\tau\ra\infty).
}
Boundary conditions at zero are set by the Wilson loop:
\eq{
R(0)=1,~~~~~P(0)=\infty,~~~~~\varPhi(0)=0.
}

The solution of equations of motion for $\varPhi$ is very simple:
\eq{
\varPhi=ij\tau.
}
It is complex, but becomes real after Wick rotation 
on the world sheet, when the internal metric has the
Minkowski signature. The solution then describes
the rotation along the
big circle of $S^5$ with constant angular momentum $J$. 

The AdS part of the minimal surface is more complicated. 
The equations of motion which follow from the action  \rf{ac} are:
\ar{
\ddot{P}-\e^{2P}\br{\dot{R}^2+R^2}&=&0,
\\*
\ddot{R}+2\dot{P}\dot{R}-R&=&0.
}
These equations admit 
two first integrals: the energy associated with translations
in $\tau$ and the dilatation charge associated with the invariance
of the action under rescaling of AdS coordinates: $R\ra\e^\omega R$,
$P\ra P-\omega$. The boundary conditions fix the values of the conserved
charges:
\ar{\label{ener}
\dot{P}^2+\e^{2P}\br{\dot{R}^2-R^2}&=&j^2,
\\*
\e^{2P}R\dot{R}-\dot{P}&=&j.
}
The first of these two equations is equivalent to Virasoro
constraints: it ensures that the induced metric is
 conformal to the flat metric. The fact that Virasoro constraints
follow from the equations of motion and boundary conditions
is a consequence of marginality
of the vertex operator.

The equations of motion can be easily integrated:
\ar{\label{mins2}
Y=\e^{-P}&=&\e^{j\tau}\left[
\sqrt{j^2+1}\tanh\br{\sqrt{j^2+1}\,\tau+\xi}-j
\right],
\non
R&=&\frac{\sqrt{j^2+1}\,\e^{j\tau}}{\cosh\br{\sqrt{j^2+1}\,\tau+\xi}}\,,
}
where
\eq{
\xi=\ln\br{\sqrt{j^2+1}+j}.
}
The solution describes a surface of revolution in $AdS_5$,
which terminates on the boundary and which has an infinite spike
that goes down to the horizon. The spike describes emission of the
external supergravity state.

There are several subtleties in evaluating the
action on the classical solution \rf{mins2}.
The action potentially
diverges on both limits of integration. The UV divergence at
$\tau=\infty$ cancels between the $AdS_5$ and the $S^5$ contributions.
Here, again, the marginality of the vertex operator
plays the crucial role. 
In particular, the boundary terms are finite:
\eq{\label{boun}
\lim_{\tau\ra\infty} \br{jP(\tau)- ij\varPhi(\tau)}
=-j\ln\br{\sqrt{j^2+1}-j}.
} 
The action also diverges at $\tau=0$. This divergence is
associated with the singularity of the metric at the boundary of
$AdS_5$.  Such kind of divergences are well-known and are
 well understood \cite{MAL,DGO}. The correct way to deal with them is
very simple: the divergence should be regularized and then
subtracted. In the present case, we do not even need to do the
subtraction, because the divergence cancels
against the normalization factor, which is given by
the classical action at $j=0$. 
Intermediate calculations still require a regularization,
which can be implemented by imposing lower bound on the integration 
variable $\tau$ at some small but finite $\ep$.

Using the conservation of energy \rf{ener}, we get for the
bulk part of the action:
\eq{\label{bulk1}
S_{\rm bulk}=\ap\, W(j)
}
with
\eq{
W(j)=\frac12\int_\ep^\infty d\tau\,\br{j^2+2\e^{2P}R^2+\dot{\varPhi}^2}
=\int_\ep^\infty d\tau\,\frac{R^2}{Y^2}\,.
}
The integrand here is a total derivative. The easiest way to see it
is to introduce
\eq{
B\equiv-\frac{\dot{R}}{R}
=
\sqrt{j^2+1}\tanh\br{\sqrt{j^2+1}\,\tau+\xi}-j.
}
The equations of motion imply that
$$
\frac{R^2}{Y^2}=\frac{\dot{B}}{B^2}.
$$
So,
$$
W(j)=\frac{1}{B(\ep)}-\frac{1}{B(\infty)}.
$$
Since
$$
B(\ep)=\ep-j\ep^2+\ldots,
$$
and 
$$
B(\infty)=\sqrt{j^2+1}-j,
$$
we get
\eq{\label{bulk2}
W(j)-W(0)=1-\sqrt{j^2+1}.
}
Collecting together all contributions, \rf{boun}, \rf{bulk1},
and \rf{bulk2}, we obtain:
\eq{\label{nn}
\cl{S}=\ap\left[1-\sqrt{j^2+1}-j\ln\br{\sqrt{j^2+1}-j}\right].
}
The OPE coefficients are determined by the classical action up
to an overall factor of order one:
\eq{
\cj=\const\cdot\exp\br{-\cl{S}}.
}
This coincides with the field-theory prediction \rf{Blim}.

\newsection{Perturbative regime}
\label{stringpert}

The two scaling limits discussed in the previous sections are related
--  the large-$J$ limit
with $J\sim \ll^{1/4}$  can be reproduced from the BMN limit,
$J\sim\ap$, by taking $j\equiv J/\ap$ small. What happens in the opposite regime
of large $j$? If $j$ were the only parameter in the problem, increasing
$j$ could be achieved either by raising $J$ or by lowering $\ll$.
In the latter case, the limit of large $j$ would correspond to the 
weak-coupling regime, which could be confronted with SYM perturbation
theory. Comparison with perturbation theory makes sense
if  large angular momentum alone can
suppress quantum fluctuations of the world sheet, without any assumptions
about the string tension. Otherwise, $\ll$ should still
be large for the semiclassical approximation to work, and then
comparison to perturbation theory makes no sense. The 
perturbative regime seems to be within the reach of the
semiclassical approximation in the
closed string sector \cite{BMN,GKP,Kristjansen:2002bb,Gross:2002su,
Constable:2002hw,Santambrogio:2002sb}. The analysis
below gives some indications that this is true for open strings
as well.

\begin{figure}[h]
\begin{center}
\epsfxsize=13cm
\epsfbox{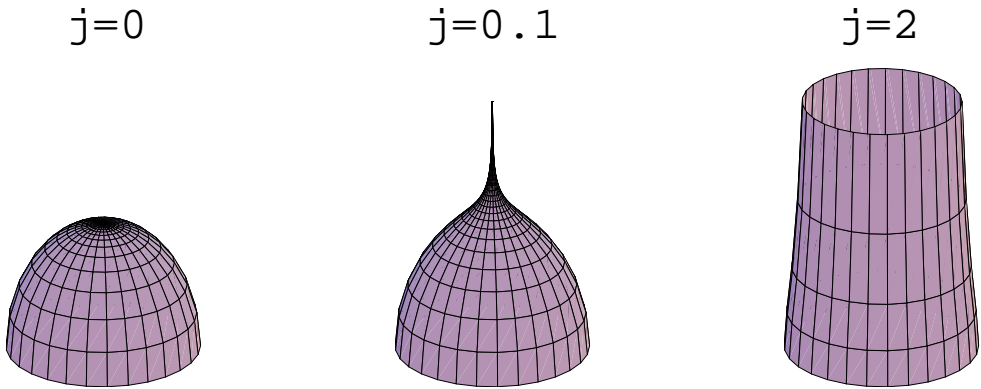}
\end{center}
\caption[x]{\small The classical string world sheet
\rf{mins2} for different values of angular momentum.}
\label{soliton}
\end{figure}

The shape of the classical world sheet \rf{mins2}  
strongly depends on the angular momentum (fig.~\ref{soliton}).
At small $j$, the
world sheet deviates very little from the solution without an operator
insertion, except for in a small
vicinity of the origin, where the world sheet degenerates into an infinite,
narrow tube that describes emission of the supergravity mode.
The width of the tube grows with $j$, and
the world sheet becomes almost a perfect cylinder at large $j$.
Indeed, the scale of variation of $R$ in \rf{mins2} is $j$ at 
$j\ra\infty$,
while the scale of variation of $Y$ is $1/j$, so $R$ is almost constant
compared to $Y$. 
This, in fact, is a general property which
is valid for any Wilson loop, not only for the  circle. 

For an arbitrary Wilson loop $W(C)$, the minimal surface 
can be parameterized by the natural parameter
$s$ on contour $C$ (such that $(dx^\mu/ds)^2=1$) and the
'time' variable $\tau$. An asymptotic solution 
at large $j$ is described by the following ansatz:
\eq{
X^\mu=x^\mu(s),~~~~~Y=\e^{-P(\tau)},~~~~~\varPhi=ij\tau.
}
The (approximate) time-independence
of $X^\mu$ is consistent with the equations of motion
in the sigma model:
\eq{
\frac{\d^2 X^\mu}{\d s^2}
+2\,\frac{\d P}{\d\tau}\,\frac{\d X^\mu}{\d\tau}
+\frac{\d^2 X^\mu}{\d \tau^2}= 0.
}
The first term is of order one at $j\ra\infty$. Assuming that
$\d P/\d\tau=O(j)$, we find that $\d X^\mu/\d\tau=O(1/j)$
and that the last term is negligible.
The time dependence of the AdS radius is determined
by the equation
\eq{
\frac{\d^2 P}{\d\tau^2}-\e^{2P}=0.
}
Imposing the boundary conditions we get:
\eq{
Y=\frac{\sinh(j\tau)}{j}.
}
This solution is universal. It is the same for any Wilson loop. 
The classical action can be readily computed:
\eq{
S_{\rm bulk}=\ap\int_\ep^\infty\frac{d\tau}{Y^2}=
-\ap\,j+\ap\,\frac{1}{\ep}\longrightarrow -J.
}
The $1/\ep$ divergence is spurious and should be subtracted \cite{MAL,DGO}.
We should also take into account the vertex operator 
evaluated on the classical solution:
a factor of
$$\br{Y(\infty)}^J\e^{-iJ\varPhi(\infty)}=(1/j)^J.$$ 
We get for the correlator:
\eq{\label{pert}
\vev{W(C)\O_J(x)}
\propto\br{\frac{\ap}{J}}^J\e^J\propto\frac{\ll^{J/2}}{J!}\,
}
for any smooth contour $C$. 

It is not hard to recognize the leading order of
SYM perturbation theory
in \rf{pert}. Indeed, the Wilson loop operator \rf{php} should
be expanded to $J$th order in scalar fields to get a non-vanishing
 correlator with $\oj$. Wilson loop is an
exponential, hence a factor of $1/J!$ will arise. The lowest-order
diagram of perturbation theory
 contains $J$ scalar propagators that connect the Wilson loop
with the local operator. Each propagator gives a factor of $\ll$.
The operator itself is proportional to $\ll^{-J/2}$, so
the correlator $\vev{W(C)\O_J(x)}$ is proportional to $\ll^{J/2}$ to the
first non-vanishing order in perturbation theory.

\newsection{Remarks}

The OPE coefficients of the circular Wilson loop
 can be computed to all orders in the SYM perturbation theory and,
for large R-charges, to all orders in the $\alpha'$ expansion in AdS
string theory. The results of these calculations completely agree in
two scaling limits, $J\sim\ll^{1/4}$ and $J\sim\ll^{1/2}$. For the
OPE coefficients, 
the interpolation between the weak-coupling and the strong-coupling
regimes can be traced on both sides
of the AdS/CFT correspondence. 

Most interestingly, the perturbative SYM regime seems to be
accessible in string theory.
The leading order
of perturbation theory is described by an approximate classical solution
in the AdS sigma model described in sec.~\ref{stringpert}. 
In principle, the equations of motion can be
solved by iterations starting from the  solution of sec.~\ref{stringpert}
as zeroth-order approximation. 
It  would be very interesting to understand if the iterative solution
of the sigma model is equivalent to ordinary planar perturbation
theory. If true, this opens an intriguing possibility
to understand how planar Feynman diagrams arise
in AdS string theory.

\subsection*{Acknowledgments}
I am grateful to J.~Minahan, M.~Staudacher and especially to
A.~Tseytlin for discussions and comments.
This work was supported by the Royal Swedish Academy of Sciences
and by STINT grant IG 2001-062
and, in part, by RFBR grant 02-02-17260 and grant
00-15-96557 for the promotion of scientific schools.

\setcounter{section}{0}
\appendix{The large-$J$ limit in field theory}
\label{bessel}

The following integral representation of the modified Bessel functions
is useful in taking the large-$J$ limit:
\eq{\label{ibb}
I_J\br{\ap}=\frac{\br{\frac{\sqrt{\ll}}{2}}^J}{\sqrt{\pi}\,\Gamma\br{J+\frac12}}
\,\int_{-1}^1 dx\,(1-x^2)^{J-1/2}\e^{\ap\,x}.
}
When $\ll\ra\infty$, $J\ra\infty$, and $J^2/\ap$ is fixed,
a convenient change of integration variable is
$t=\ap(x+1)$:
\eq{\label{intbess}
I_J\br{\ap}=\frac{\ll^{-1/4}\e^{\ap}}{\sqrt{2\pi}\Gamma\br{J+\frac12}}
\,\int_{0}^{2\ap}dt\,\e^{- t}t^{J-1/2}\br{1-\frac{t}{2\ap}}^{J-1/2}.
}
The strong-coupling expansion is generated
by expanding the last factor in $t/2\ap$.
At large $J$, the integral is dominated by $t\sim J$, 
 and the expansion breaks down. 
Instead of expanding the last factor, we should replace
it by
an exponential:
$$
\br{1-\frac{t}{2\ap}}^{J-1/2}
=\br{1-\frac{t(J-1/2)}{2\ap(J-1/2)}}^{J-1/2}
\approx\exp\left[-\frac{t(J-1/2)}{2\ap}\right].
$$
The integration over $t$ is then trivial:
$$
I_J\br{\ap}\approx\frac{\ll^{-1/4}\e^{\ap}}{\sqrt{2\pi}}\,
\br{1-\frac{J-1/2}{2\ap}}^{-(J-1/2)}
\approx\frac{\ll^{-1/4}\e^{\ap}}{\sqrt{2\pi}}\,
\exp\br{-\frac{J^2}{2\ap}}.
$$
Dividing by the large-$\ll$ asymptotics
of $I_1\br{\ap}$,
$$
I_1\br{\ap}\approx\frac{\ll^{-1/4}\e^{\ap}}{\sqrt{2\pi}}\,,
$$
we get \rf{dsl}.

In the BMN limit: $\ll\ra\infty$, $J=j\ap$, and $j$ is finite.
The integral \rf{ibb} can be computed by the saddle-point method
in this limit. The integral is dominated by the maximum of
$$
{\cal S}(x)=\ap\left[x+j\ln\br{1-x^2}\right],
$$
which is reached
at $x\equiv x_0=\sqrt{j^2+1}-j$, and
$$
{\cal S}(x_0)=\ap\left[\sqrt{j^2+1}-j+\ln\br{2j\br{\sqrt{j^2+1}-j}}\right].
$$
The overall factor in \rf{ibb} also has a well-defined BMN limit:
$$
\frac{\br{\frac{\ll}{2}}^J}{\Gamma\br{J+\frac12}}
\approx\const\cdot\exp\left[\ap\,j(1-\ln 2j)\right].
$$
Consequently,
\eq{
I_J\br{\ap}\approx\const\cdot\exp\left\{\ap\left[\sqrt{j^2+1}+
j\ln\br{\sqrt{j^2+1}-j}\right]\right\}.
}
Dividing by the normalization factor $I_1\br{\ap}$,
we get \rf{Blim}.

\appendix{Green's functions}
\label{greens}

The classical string world sheet for the circular loop
has a geometry of $AdS_2$. The metric \rf{metr} can be put
to a more familiar form by an inversion:
\eq{
x^\mu\lra\frac{(x+x_0)^\mu}{(x+x_0)^2+z^2},~~~~~
z\lra\frac{z}{(x+x_0)^2+z^2},
}
with $x_0=(1,0,0,0)$. This transformations maps a hemi-sphere
\rf{mins} onto a half-plane $x^\mu=(1/2,u,0,0)$, $-\infty<u<\infty$,
$0<z<\infty$, which is an $AdS_2$ subspace of $AdS_5$ with the
metric $ds^2=(dz^2+du^2)/z^2$.

The scalar propagators in $AdS_2$ are well known 
\cite{Fronsdal:ew,Inami:1985wu,Burgess:ti}. 
The general result for a scalar field  in $AdS_{(d+1)}$ with 
$m^2=\Delta(\Delta-d)$ is \cite{D'Hoker:2002aw}:
\eq{
G_{m^2}(z,u;z',u')=
\frac{\Gamma(\Delta)}{2^\Delta\pi^{d/2}\Gamma(\Delta-d/2)(2\Delta-d)}\,
\xi^\Delta F\br{\frac{\Delta}{2},\frac{\Delta+1}{2};\Delta-\frac{d}{2}+1;
\xi^2},
}
where
\eq{
\xi=\frac{2zz'}{z^2+{z'}^2+(u-u')^2}.
}
Specifying to $d=1$ and to $m^2=0,2$; we get:
\ar{
G_0(\xi)&=&\frac{1}{4\pi}\,\ln\br{\frac{1+\xi}{1-\xi}}~~~~~(\Delta=1),
\non
G_2(\xi)&=&\frac{1}{4\pi\xi}\,\ln\br{\frac{1+\xi}{1-\xi}}
-\frac{1}{2\pi}~~~~~(\Delta=2).
}
In the limit of coinciding points, $\xi\ra 1$. Therefore,
\eq{
A=\lim_{\xi\ra 1}\br{G_0(\xi)-G_2(\xi)}=\frac{1}{2\pi}.
}

\end{document}